\documentclass{article}

\usepackage{arxiv}

\usepackage[utf8]{inputenc} 
\usepackage[T1]{fontenc}    
\usepackage{hyperref}       
\usepackage{url}            
\usepackage{booktabs}       
\usepackage{amsfonts}       
\usepackage{amsmath}
\usepackage{nicefrac}       
\usepackage{microtype}      
\usepackage{cleveref}       
\usepackage{lipsum}         
\usepackage{graphicx}
\usepackage{natbib}
\usepackage{doi}

\title{Frequency Chirping of Electromagnetic Ion Cyclotron Waves in Earth's Magnetosphere}

\author{
	\href{https://orcid.org/0000-0002-0154-0725}{\includegraphics[scale=0.06]{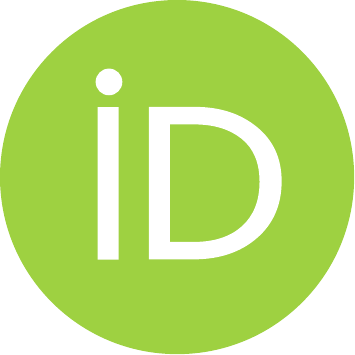}\hspace{1mm}Zeyu An} \\
	Department of Geophysics and Planetary Sciences\\
	University of Science and Technology of China\\
	Hefei, China \\
	\And
	\href{https://orcid.org/0000-0002-4676-8133}{\includegraphics[scale=0.06]{orcid.pdf}\hspace{1mm}Xin Tao
	\thanks{Correspondence to: \texttt{xtao@ustc.edu.cn}.}} \\
	Department of Geophysics and Planetary Sciences\\
	University of Science and Technology of China\\
	Hefei, China \\
	\And
	\href{https://orcid.org/0000-0002-9270-4704}{\includegraphics[scale=0.06]{orcid.pdf}\hspace{1mm}Fulvio Zonca} \\
	Center for Nonlinear Plasma Science and C.R. ENEA Frascati \\
	C.P. 65, 00044 Frascati, Italy \\
	\And
	\href{https://orcid.org/0000-0002-4180-588X}{\includegraphics[scale=0.06]{orcid.pdf}\hspace{1mm}Liu Chen} \\
	Department of Physics and Astronomy \\
	University of California, Irvine \\
	CA 92697, USA \\
}

\renewcommand{\shorttitle}{EMIC Wave Frequency Chirping}

\hypersetup{
pdftitle={Frequency Chirping of Electromagnetic Ion Cyclotron Waves in Earth's Magnetosphere},
pdfsubject={geophysics},
pdfauthor={Zeyu An, Xin Tao, Fulvio Zonca, Liu Chen},
}

\begin{document}
\maketitle

\begin{abstract}
Electromagnetic ion cyclotron waves are known to exhibit frequency chirping, contributing to the rapid scattering and acceleration of energetic particles. However, the physical mechanism of chirping remains elusive. Here, we propose a new model to explain the chirping and provide direct observational evidence for validation. Our results relate the frequency chirping of the wave to both the wave amplitude and magnetic field inhomogeneity for the first time. The general applicability of the model's underlying principle opens a new path toward understanding the frequency chirping of other waves. 
\end{abstract}


\section{Introduction}
\label{sec:intro}
Frequency chirping, a rapid change in frequency over time, has been observed in several wave modes in Earth's magnetosphere and laboratory plasmas. In the magnetosphere, examples of frequency chirping waves include whistler mode chorus waves \citep{Tsurutani1974}, electromagnetic ion cyclotron (EMIC) waves \citep{Mursula2007,Pickett2010}, magnetosonic waves \citep{Boardsen2014,Fu2014}, and electron cyclotron harmonic waves \citep{Shen2021,Teng2021}. Similarly, in fusion plasmas, Alfv\'{e}n modes have been observed to exhibit frequency chirping on different time scales \citep{McGuire1983,Heidbrink1994,Heidbrink2008}. These chirping waves generally consist of discrete packets that are narrowband and quasi-coherent \citep{Tsurutani2009}. Interaction with these chirping modes often results in the rapid transport of energetic particles in phase space \citep{McGuire1983,Chen2007,Artemyev2016,Zhang2018,Zonca2005,White1983}, affecting space weather or deteriorating particle confinement in fusion devices. Therefore, there is significant research interest in understanding the physical mechanism of frequency chirping in both communities of space and laboratory plasmas.

In this Letter, we focus on the frequency chirping of EMIC waves in the magnetosphere, which has been studied only preliminarily. The only theoretical model \citep{Omura2010} of EMIC wave chirping we know of is based on nonlinear wave-particle interactions, which produces a chirping rate proportional to EMIC wave amplitude. Here, we propose a new model of EMIC wave chirping based on nonlinear wave-particle interaction theories of EMIC waves and the principle of the recently proposed ``Trap-Release-Amplify'' (TaRA) model \citep{Tao2021} of whistler mode chorus waves. Our new model relates the chirping rate of EMIC waves not only to wave amplitude via nonlinear wave-particle interactions but also to the background magnetic field inhomogeneity for the first time, providing a more comprehensive understanding of the underlying mechanisms of EMIC wave chirping. Additionally, we provide a direct quantitative comparison between theoretical chirping rates and observations. The comparison shows good agreement, suggesting the validity of our new model and that the same underlying principle of chirping mechanisms exists between EMIC waves and chorus waves. Our results, therefore, not only improve the understanding of the chirping of EMIC waves but also open new perspectives for further investigating the chirping mechanism of other wave modes in both space and laboratory plasmas.

\begin{figure}
\includegraphics[width=\textwidth]{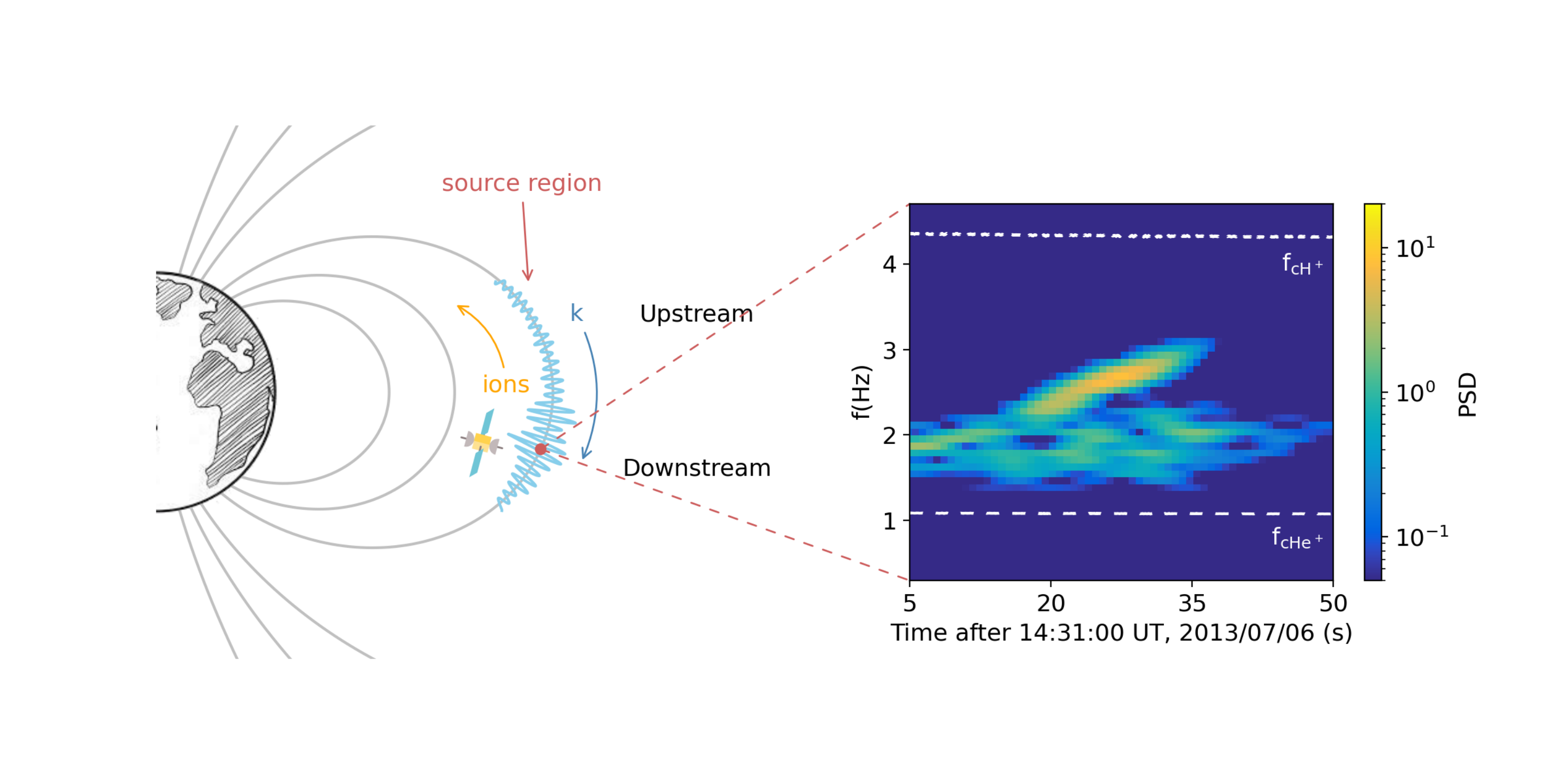}
\caption{\label{fig:TaRA}An illustration of the  EMIC wave chirping model. As indicated by the satellite icon, chirping elements are usually observed downstream of the equator, where they have been fully amplified through nonlinear wave-particle interactions. The rising-tone EMIC wave element shown in the right panel was observed by RBSP-A satellite at 14:31 UT, 2013/07/06.}
\end{figure}

\section{Model}
\label{sec:TaRA}
We illustrate our chirping model of EMIC waves in Figure \ref{fig:TaRA}. Each element of the chirping EMIC wave is represented by a narrowband quasi-coherent wave packet, with a frequency that increases nearly continuously from $\omega_{\min}$ to $\omega_{\max}$. We further adopt the basic principle of the TaRA model of chorus and divide the generation region of EMIC waves into upstream and downstream regions separated by the equator. As the generated part of the EMIC wave packet propagates downstream of the equator, it constantly encounters fresh ions moving in the opposite direction and satisfying the cyclotron resonance condition. If the wave is quasi-coherent with a large enough amplitude, these ions get phase-trapped by the wave packet and form a localized phase space hole, which has been demonstrated by computer simulations in \citet{Shoji2011,Shoji2013}. As the ions travel across the wave packet, they will eventually get released in the upstream region where the wave amplitude is small. However, these released ions still retain certain phase correlation, resulting in a current and selective amplification of new emissions from the nearly continuous spectrum of background EMIC waves. The condition for the selective amplification is mathematically expressed using the phase-locking condition $\mathrm{d}^2 \zeta / \mathrm{d} t^2 = 0$. Here, $\zeta$ is the wave-particle interaction phase angle between ion perpendicular velocity $\boldsymbol{v}_{\perp}$ with respect to the background magnetic field, and the wave magnetic field $\delta \boldsymbol{B}$. Such a selection rule guarantees maximum power transfer from ions to waves by holding the resonance condition $\mathrm{d} \zeta / \mathrm{d} t = 0$ for the longest time possible. Note that fresh ions are continuously entrapped in the downstream and detrapped in the upstream, allowing frequency chirping from $\omega_{\min}$ to $\omega_{\max}$.

Based on the above chirping model, two different chirping rates can then be derived. From the nonlinear wave-particle interaction theories of parallel propagating EMIC waves, the second-order time derivative of the wave-particle interaction phase angle $\zeta$ could be written as \citep{Omura2010}

\begin{equation}
\label{eq:2}  
\frac{\mathrm{d}^2 \zeta}{\mathrm{d} t^2} = \omega_{tr}^2 \sin \zeta + (R_1 + R_2),
\end{equation}

where

\begin{equation}
\label{eq:3}
R_1 = \left( 1 - \frac{v_r}{v_g} \right)^2 \frac{\partial \omega}{\partial t},
\end{equation}

\begin{equation}
\label{eq:4}
R_2 = \left[ \left( \frac{v_{\perp}^2}{2 v_p} + \frac{v_r^2}{v_g} \right) \frac{\omega}{\Omega_i} - v_r \right] \frac{\partial \Omega_i}{\partial s},
\end{equation}

assuming constant cold plasma density along the field line. Here $R_1$ and $R_2$ represent the effects of frequency chirping and background magnetic field nonuniformity, respectively, and $s$ is the distance along a field line from the equator. The parameter $R$, widely used by other studies \citep{Vomvoridis1979,Nunn1974}, equals $(R_1 + R_2) / \omega_{tr}^2$. In Equation \ref{eq:2} above, $\omega_{tr} = \sqrt{k v_{\perp} \Omega_w}$ is the phase-trapping frequency, with $k$ the wavenumber and $\Omega_w \equiv q_i \delta B / m_i c$. Here $q_i$ is the ion charge, $m_i$ is the ion mass and $c$ is the speed of light in vacuum. In Equations \ref{eq:3} and \ref{eq:4}, $v_r = (\omega - \Omega_i) / k$ is the cyclotron resonance velocity, $v_g = \partial \omega / \partial k$ is the wave group velocity, $v_p = \omega/k$ is wave phase velocity, and $\Omega_i \equiv q_i B / m_i c$ is the non-relativistic ion cyclotron frequency. 

\begin{figure}
\includegraphics[width=\textwidth]{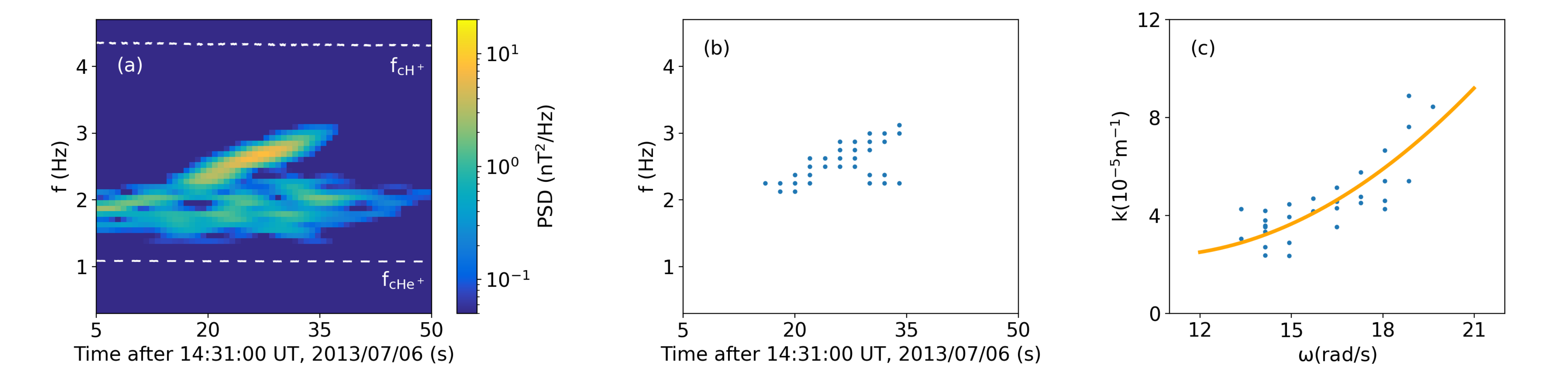}
\caption{\label{fig:wave}A demonstration of the wavenumber estimation method. (a) The spectrogram of the chirping element, Event 10 of Table \ref{tab:cases}. (b) Points in the spectrogram which satisfy all selection rules described in the text. (c) Wavenumber $k$ solved with the SVD method (the blue dots) and a quadratic fit (the orange curve) of them as a function of frequency $\omega$.}
\end{figure}

The first chirping rate is deducted from nonlinear wave-particle interaction theories and is same as the one obtained by the previous model of EMIC wave chirping. At the equator, $R_2 \propto \partial \Omega_i / \partial z \sim 0$. We then arrive at the relation between chirping rates and wave amplitudes as in earlier studies \citep{Vomvoridis1982,Omura2008,Omura2010, Zonca2022}:

\begin{equation}
\label{eq:6}
\frac{\partial \omega}{\partial t} = R \left( 1 - \frac{v_r}{v_g}  \right)^{-2} \omega_{tr}^2 \propto \delta B.
\end{equation}

Effective power transfer is found to occur for $R$ to be about $0.5$ \citep{Vomvoridis1982, Omura2008, Zonca2021}. This chirping rate has been obtained before, and for convenience, it will be called the nonlinear chirping rate. 

Applying the principle of the TaRA model of chorus to EMIC waves allows us to obtain a new expression of the chirping rate. In the source region, phase-trapped ions are released to amplify new waves from the pre-existing thermal noise spectrum, which means that $\delta B$ is small or $\omega_{tr}^2$ is negligible compared with $R_2$ in Equation \ref{eq:2}. The phase-locking condition in the upstream region is then reduced to

\begin{equation}
\frac{\mathrm{d}^2 \zeta}{\mathrm{d} t^2} \approx R_1 + R_2 = 0,
\end{equation}

leading to

\begin{equation}
\label{eq:5}
\frac{\partial \omega}{\partial t} = - \left( 1 - \frac{v_r}{v_g} \right)^{-2} \left[  \left( \frac{v_{\perp}^2}{2 v_p} + \frac{v_r^2}{v_g}  \right) \frac{\omega}{\Omega_i} - v_r\right] \frac{\partial \Omega_i}{\partial s}.
\end{equation}

This equation demonstrates the dependence of chirping rate on background magnetic field inhomogeneity and is expected to be evaluated at the wave source region upstream of the equator. Denoting this location as $s_0$, an order of magnitude estimate of $s_0$ is that $s_0 = (2 \pi | v_r | / \xi \Omega_{i0})^{1/3}$ \citep{Helliwell1967}, corresponding to a shift of $\pi$ radian of $\zeta$ from the equator assuming linear motion of ions. Here we approximate the background magnetic field $B$ near the equator as $B=B_0(1+\xi s^2)$, and $\xi$ is a measure of the magnetic field inhomogeneity.  

So far, two chirping rate expressions have been derived: Equation \ref{eq:6} at the equator and Equation \ref{eq:5} in the source region.  The new model of EMIC wave chirping, therefore, predicts that the chirping rate of EMIC waves is not only related to wave amplitude at the equator, but also to the background magnetic inhomogeneity. We must emphasize that, although the two equations look different and have different origins of physical mechanisms, they are consistent and both are correct at different stages of wave excitation. 

\section{Observation}
\label{sec:results}
We now provide a direct comparison with observation to test the above two EMIC wave chirping rates using all previously published events we are aware of. Different from whistler mode chorus waves, there are not many observed events of chirping EMIC waves. The only selection rule we apply is that the magnetic latitude of the event should be less than $5^{\circ}$ for Equation \ref{eq:6} or $10^{\circ}$ for Equation \ref{eq:5} to avoid being too far away from the equatorial source region. We select events closer to the source region for Equation \ref{eq:6} because the wave amplitude could vary significantly as the wave propagates from the equator to high latitudes. In total, we have twelve observed chirping EMIC wave elements for Equation \ref{eq:5} and nine for Equation \ref{eq:6}, all of which are in-situ observations of EMIC waves with clear frequency chirping element. These observations were made using different spacecraft including Cluster \citep{Escoubet1997}, Van Allen Probes \citep{Mauk2013}, and Time History of Events and Macroscale Interactions during Substorms (THEMIS, \citet{Angelopoulos2008}), during different time periods and at a wide range of radial distances. Therefore, these selected events should be a good representation of EMIC wave chirping elements in the magnetosphere. The details of these observed events are summarized in Table \ref{tab:cases}.

\begin{table}
\caption{\label{tab:cases}Detailed information of the 12 observational events. For each event, the time of occurrence is a 10-minute-long vicinity of the interested chirping element.}
\resizebox{\linewidth}{!}{
\begin{tabular}{cccccc}
\toprule
&Time of occurrence (UT)&L-shell&MLAT (deg)&Observed by&First reported by\\
\midrule
1&07:55-08:05, 2002/03/30&4.45&-3.4&Cluster 4&Pickett et al. \citep{Pickett2010}\\
2&00:10-00:20, 2003/03/27&4.17&-1.78&Cluster 4&Grison et al. \citep{Grison2013}\\
3&09:05-09:15, 2008/04/16&7.2&-8.82&THEMIS A&Nakamura, Omura, Shoji et al. \citep{Nakamura2015}\\
4&17:40-17:50, 2008/10/11&6.93&0.2&THEMIS A&Nakamura, Omura, Shoji et al. \citep{Nakamura2015}\\
5&18:30-18:40, 2010/06/25&8.03&0.08&THEMIS A&Nakamura, Omura, Shoji et al. \citep{Nakamura2015}\\
6&17:35-17:45, 2010/07/11&7.82&3.68&THEMIS E&Nakamura, Omura, Machida et al. \citep{Nakamura2014}\\
7&14:20-14:30, 2010/09/09&8.33&4.83&THEMIS D&Nakamura, Omura, Machida et al. \citep{Nakamura2014}\\
8&13:20-13:30, 2010/09/24&8.27&4.8&THEMIS D&Nakamura, Omura, Shoji et al. \citep{Nakamura2015}\\
9&08:35-08:45, 2011/09/25&8.88&-2.89&THEMIS A&Nakamura, Omura, Shoji et al. \citep{Nakamura2015}\\
10&14:25-14:35, 2013/07/06&4.05&2.45&RBSP-A&Chen et al. \citep{Chen2019}\\
11&10:15-10:25, 2015/12/22&5.4&-8.02&RBSP-A&Sigsbee et al. \citep{Sigsbee2020}\\
12&12:15-12:25, 2017/09/07&5.68&5.63&RBSP-A&Zhu et al. \citep{Zhu2020}\\
\bottomrule
\end{tabular}}
\end{table}

Before calculating theoretical chirping rates, we describe the method we use to estimate wavenumber $k$. Although it is typical to obtain the wavenumber from cold plasma dispersion relation, this estimate could be very inaccurate for frequencies near ion cyclotron frequencies. Furthermore, the heavy ion ($\mathrm{He^+}$ and $\mathrm{O^+}$) densities are sometimes difficult to determine or even missing in satellite data. Using statistical heavy ion compositions is inappropriate here, as different events might correspond to very different cold plasma parameters. To avoid this problem, we adapt the wavenumber analysis method proposed by \citet{Chen2019}, which allows the determination of $k$ as a function of $\omega$ directly from the observed waveform. The method involves three steps for a chirping element like the one shown in Figure \ref{fig:wave}(a). First, we determine the evolution of wave frequency with time from the wave spectrogram. In this step, we only select data points with large enough power spectral density and a small enough wave normal angle. Specifically, we require that the magnetic power spectral density $\mathrm{PSD} > 0.01 \mathrm{PSD}_{\max}$, where $\mathrm{PSD}_{\max}$ is the maximum $\mathrm{PSD}$ in the chirping element, and wave normal angle $\Psi < 45^\circ$. The selected data points from the spectrogram are shown in Figure \ref{fig:wave}(b). Second, we use the singular value decomposition (SVD) method \citep{Santolik2003} to solve for $\boldsymbol{k}$ at the selected points, using five components of the electromagnetic fields ($B_x$, $B_y$, $B_z$, $E_y$, and $E_z$). The wavenumbers of the selected data points are plotted against $\omega$ in Figure \ref{fig:wave}(c). Finally, we fit the data with a quadratic function of $\omega$: $k = a \omega^2 + b \omega + c$ to obtain a single wavenumber as a function of frequency. When estimating the chirping rate of a given element, we use $\omega_{mid} = (\omega_{\min} + \omega_{\max}) / 2$ and the corresponding $k(\omega_{mid})$ to represent an average rate for the whole element. 

\begin{figure}
\includegraphics[width=\textwidth]{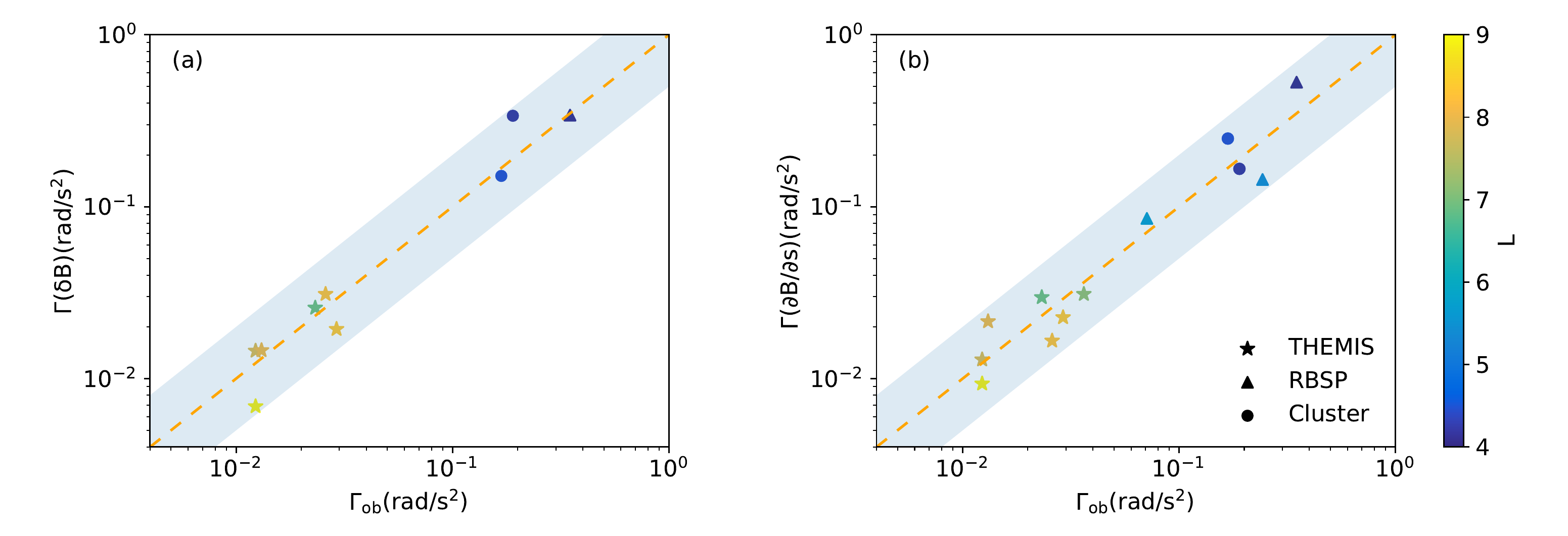}
\caption{\label{fig:chirping}Comparisons between observational ($\Gamma_{ob}$) and theoretical chirping rates calculated with (a) Equation \ref{eq:6} ($\Gamma (\delta B)$) and (b)  Equation \ref{eq:5} ($\Gamma (\partial B / \partial s)$). Different markers and colors categorize the data source and L-shell value of each case, respectively. The shaded region is confined by two lines $\Gamma (\partial B / \partial s)$ (or $\Gamma (\delta B)$) $= 0.5 \Gamma_{ob}$ and $2 \Gamma_{ob}$, while the orange dashed line denotes $\Gamma (\partial B / \partial s)$ (or $\Gamma (\delta B)$) $= \Gamma_{ob}$.}
\end{figure}

Figure \ref{fig:chirping} shows comparisons between observational ($\Gamma_{ob}$) and theoretical chirping rates. Figure \ref{fig:chirping}(a) shows the comparison for $\Gamma (\delta B)$, Equation \ref{eq:6}, and Figure \ref{fig:chirping}(b) for $\Gamma (\partial B / \partial s)$, Equation \ref{eq:5}. When estimating theoretical chirping rates, we calculate $\delta B$ by integrating over the wave power spectral density. Both the equatorial value $B_0$ and the inhomogeneity factor $\xi$ of the background magnetic field are derived by fitting the data using the T89 magnetic field model \citep{Tsyganenko1989} with the parabolic function $B=B_0(1+\xi s^2)$. The perpendicular velocity $v_{\perp}$ is estimated from resonant velocity $v_r$ and a pitch angle of $70^{\circ}$, following that nonlinear phase-trapping occurs most easily for pitch angle between $65^{\circ}$ and $75^{\circ}$ \citep{Inan1978}.  For observational chirping rate, we simply use $\Gamma_{ob} = (\omega_{\max}-\omega_{\min})/\Delta t$ with $\Delta t$ the duration of the element. Each point represents an event in Table \ref{tab:cases}, with the corresponding satellite missions and L-shell values categorized by different markers and colors, respectively. All data points are located inside the shaded region, indicating that the differences between observation and both theoretical chirping rates are less than a factor of two. Correspondingly, the direct comparison presented in Figure \ref{fig:chirping} demonstrates very good agreement between the new model predictions and observation. Furthermore, similar to chorus waves \citep{Tao2012, Tao2014c, Xie2021}, we can see a decrease of chirping rate as $L$ increases because of smaller magnetic field inhomogeneity.

The discrepancies between observations and theoretical predictions come mainly from uncertainties in some parameters, such as wave amplitude ($\delta B$) and source location ($s_0$). Although we select events with $|\lambda| < 5^{\circ}$ when calculating $\delta B$, the wave amplitude might still be different from its equatorial value. Furthermore, the equation $s_0 = (2 \pi \vert v_r \vert / \xi \Omega_{i0})^{1/3}$ we use to locate the source region is only an estimation to the lowest order by assuming adiabatic motions of ions, while to be exact, the release location needs to be calculated by taking nonlinear wave particle interactions into consideration. Other potential causes of estimation errors include the assumption of constant cold plasma density along the field line, and the fitting of wave $k(\omega)$. In general, taking these possible sources of uncertainties into consideration, we have reached good agreement between observations and two different yet consistent theoretical chirping rate estimates. Therefore, we conclude that the model proposed in Section \ref{sec:TaRA} could well explain the chirping of EMIC waves. 

\section{Summary}
\label{sec:summary}
In this Letter, we present a new model of EMIC wave chirping and directly compare it with in-situ satellite observations. The two theoretical predictions of chirping rates exhibit good agreement with observations, indicating that both rates are manifestations of the same underlying process at different stages of wave packet propagation. The new model not only relates EMIC wave chirping to wave amplitude at the equator but also to background magnetic field inhomogeneity, which is a novel contribution. Furthermore, our comparison confirms that the principle of the TaRA model, initially developed for whistler mode chorus, is equally applicable to EMIC wave chirping. We note that some components of the TaRA model, such as phase-locking and detrapping, are also shared by Alfv\'{e}n wave chirping in fusion plasmas from a recent study \citep{Wang2022}. Combining these findings suggests the possibility of a unified theoretical framework to explain the nonlinear chirping of different waves in various plasma environments \citep{Chen2016,Zonca2022}.

\section{Acknowledgments}
We acknowledge the Cluster, THEMIS and RBSP teams for providing the satellite data. The data used by this study can be found at the corresponding data archive (Cluster, https://csa.esac.esa.int; THEMIS: http://themis.ssl.berkeley.edu; Van Allen Probes:  http://www.space.umn.edu/rbspefw-data and https://emfisis.physics.uiowa.edu/data/index.). This work was supported by the B-type Strategic Priority Program of the Chinese Academy of Sciences, Grant No. XDB41000000, NSFC grants (42174182 and 11235009), Euratom Research and Training Programme (Grant Agreement No 101052200 — EUROfusion). Views and opinions expressed are however those of the author(s) only and do not necessarily reflect those of the European Union or the European Commission.

\bibliographystyle{unsrtnat}

\providecommand{\noopsort}[1]{}\providecommand{\singleletter}[1]{#1}%

\end{document}